\documentclass[aps,prl,twocolumn,floatfix]{revtex4}
\usepackage{graphicx}
\usepackage{dcolumn}
\usepackage{latexsym}
\usepackage{amssymb}
\usepackage{amsfonts}
\usepackage{amsmath}
\usepackage{fancybox}
\usepackage[binary,squaren]{SIunits}
\usepackage{colordvi}
\newcommand{\bg}[1]{\mbox{\boldmath$#1$\unboldmath}} 

\begin{document}
\title{Collective Spin-Hall Effect for Electron-Hole Gratings}
\author{Ka Shen and G. Vignale}
\affiliation{Department of Physics and Astronomy, University of
Missouri, Columbia, Missouri 65211, USA} 
\date{\today}
\begin{abstract}
We show that an electric field  parallel to the wavefronts 
of an electron-hole grating in a GaAs quantum well generates, via the electronic spin Hall effect, a spin grating of the same wave vector and with an amplitude that can exceed  1\% of the amplitude of the initial density grating.  We refer to this phenomenon as ``collective spin Hall effect".  A detailed study of the coupled-spin charge dynamics for quantum wells grown in different directions reveals rich features in the time evolution of the induced spin density, including the possibility of generating a helical spin grating.

\end{abstract}
\maketitle
The spin Hall effect (SHE), i.e., the generation of a transverse spin
current from a charge current and vice versa, has attracted much attention in the past
decade~\cite{Dyakonov71,Hirsch99,Zhang00,Murakami03,Sinova04,kato04,Murakami05,Wunderlich05,Engel07,Hankiewicz09}, and has now become one of the standard tools for the generation and detection of spin currents in magneto-electronic devices~\cite{Zutic04,Fabian07,Awschalom07,Wu10,Liu11}. Theoretically, both intrinsic and extrinsic mechanisms have been shown to contribute to the SHE in semiconductors.
While the intrinsic mechanism originates from the spin-orbit
coupling (SOC) in the band structure~\cite{Murakami03, Sinova04}, the extrinsic one 
results from the SOC with  impurities~\cite{Dyakonov71,Hirsch99,Zhang00}. 
Experimentally, the first evidence of SHE in 
semiconductors was the observation of a spin accumulation at the edges
of n-doped GaAs~\cite{kato04}. This is clearly a single-particle effect taking place in a macroscopically homogeneous sample.
Recently, Anderson et al.~\cite{Anderson10} have proposed an
interesting collective manifestation of the SHE in a periodically modulated electron gas. 
They suggested that an optically induced spin density
wave (transient spin grating~\cite{Cameron96,weber_nature_2005,Yang12}) in a two-dimensional electron gas could be partially converted into a density wave when an electric
field perpendicular to the grating wave vector is applied.

There are some difficulties with the implementation of this idea. 
First of all, the electric field due to the induced
charge density, when properly taken into account,  effectively prevents the accumulation of  charge.   
Second, the SOC considered in that work comes solely from band structure (i.e., it is purely ``intrinsic") and, for this reason, the  spin-charge coupling is found to be of third order in the, presumably small, strength of the SOC. 

\begin{figure}[]
\includegraphics[width=6cm]{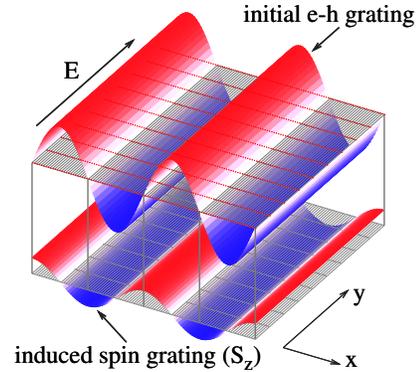}
\caption{(Color online) Collective spin Hall density profile induced by 
 a transverse electric field ($\mathbf E$) in a periodically modulated electron/hole
gas.} 
\label{model}
\end{figure}

In the present work, we re-examine  the coupled spin-density transport in a
periodically modulated electron gas in a novel set-up which is free of the above-mentioned difficulties.  
Differing from Ref.\,\onlinecite{Anderson10}, we start from an electrically neutral electron-hole grating (uniform spin density) in an $n$-type semiconductor quantum well  and show that an electric field parallel to the wavefronts of the grating generates, via  SHE,  a periodic spin modulation of the same wave vector as the initial electron-hole grating (see Fig.\,\ref{model}).
Since any local charge imbalance is screened quickly by the background
electrons, we can safely assume that the system remains charge-neutral
throughout its evolution and, in particular, no additional electric field is created.
Furthermore, going beyond the treatment of Ref.\,\onlinecite{Anderson10}, we
include not only the intrinsic but also the extrinsic SHE.  
We confirm that the spin-density
coupling due to the intrinsic SHE is an effect of third-order in the
SOC strength~\cite{footnote0},
which is consistent with previous
works~\cite{Burkov04,Galitski07,Mishchenko04,Bernevig06}. 
However, we also find that the dominant extrinsic spin Hall mechanism,  skew
scattering~\cite{Smit58}, leads to
an inhomogenous spin-charge coupling that is of  {\it first order} in the SOC strength. 

In the light of this analysis, the chances for the observation of the collective SHE appear much better than previously thought.  For electron-hole density gratings induced by optical means in an $n$-type GaAs quantum well, we predict the amplitude of the induced spin grating to be larger than 1\% of the amplitude of the original grating at an electric field of $\sim 10^5$ V/m.  These numbers are within the reach of contemporary experimental techniques~\cite{Cameron96,weber_nature_2005,Yang12}.

{\it Theoretical framework -- }
Up to the linear order in momentum, the Hamiltonian of  electrons in an $n$-GaAs QW  can be written as  
\begin{equation}
H_{0}=\frac{k^{2}}{2m_e}+\frac{1}{m}\mathbf{k} \cdot{\bf A},
\label{hamiltonian}
\end{equation}
where ${\bf A}=m(\lambda_1\sigma_{y}+\gamma_y\sigma_z,\lambda_2\sigma_{x}-\gamma_x\sigma_z)$ is the spin-dependent vector potential that describes SOC. 
Specifically, if $\alpha$ and $\beta$ denote the Rashba~\cite{Rashba60,Rashba84} and 
Dresselhaus coefficients~\cite{Dresselhaus55}, we have $\lambda_1=\beta+\alpha$,
$\lambda_2=\beta-\alpha$ and $\gamma_i=\frac{\lambda_{e}^2}{4} eE_i$
in  a (001) QW (the $x$ and $y$ axes are in the [110] and [$\bar 1$10]
directions, forming  $45^\degree$ angles with the cubic axes), while
$\lambda_1=\alpha$, $\lambda_2=-\alpha$, $\gamma_x=\frac{\lambda_{e}^2}{4} eE_x$ and $\gamma_y=
\frac{\lambda_{e}^2}{4}eE_y-\beta$ in a (110) QW .  The terms containing the effective
Compton wavelength $\lambda_{e}$ ($\sim 4.6$~\AA~in GaAs) describe the
SOC from the applied electric field. 
The Hamiltonian for the (heavy) holes has a similar form with a different effective mass $m_h$.   In this case, however, we assume that the spin polarization is quenched, due to strong spin-orbit interaction in the valence band, on a time scale that is  shorter than that of the diffusion process.  For this reason, no spin-dependent terms are included for the holes.

Our analysis is based on the quantum kinetic equation for the  density matrix 
 $\rho_{\bf  k}({\bf r})$ of electrons~\cite{Mishchenko04,Bernevig06,Wu10}:
\begin{equation}
\partial_{t}\rho_{\mathbf{k}}+\frac{1}{2}\{\nabla_{\mathbf{k}}H_{0},\tilde\nabla_{\mathbf{r}}\rho_{\mathbf{k}}\}
+i[H_{0},\rho_{\mathbf{k}}]=\partial_{t}\rho_{\mathbf{k}}|_{\text{scat}},
\label{eq_ksbe}\end{equation} 
where $\tilde \nabla_{\mathbf{r}}=\nabla_{\bf r}+e{\bf E}\partial_{\epsilon_k}$. In the relaxation time approximation, the scattering term on the right-hand side is 
\begin{eqnarray}
  \partial_{t}\rho_{\mathbf{k}}|_{\text{scat}}  &=&
  -\frac{\rho_{\mathbf{k}}}{\tau}+\frac{\rho_{k}}
  {\tau}+\frac{1}{2m\tau}\{\mathbf{k}\cdot\mathbf{A},\partial_{\epsilon_{k}}\rho_{k}\}\nonumber\\
  &&\mbox{}-\frac{1}{2}\alpha_{ss} \sum_{|\mathbf{k}'|=|\mathbf k|}
\{\mathbf{k}\times\mathbf{k}'\cdot\bg\sigma,\rho_{\mathbf{k}'}\},
  \label{RTA}
\end{eqnarray}
where  $\rho_k=\langle \rho_{\bf k}\rangle$ is  the momentum-space  angular average of the density matrix.
The last term on the right-hand side of Eq.~(\ref{RTA})  is the skew scattering term~\cite{Cheng08,Raimondi12} with the coefficient
$\alpha_{ss}=\frac{\hbar}{8\pi m}n_{i}\lambda_{e}^2 \left(\frac{mu_i}{\hbar^2}\right)^3 $, where $n_i$ and $u_i$ are the density and the scattering potential of impurities,
respectively~\cite{footnotex}.  The third term on the right hand side of
Eq.\,(\ref{RTA}), which effectively amounts to shifting the argument
of $\rho_{\bf k}$ from ${\bf k}$ to ${\bf k}+{\bf A}$,    is
critically important to ensure the vanishing of the 
spin-charge coupling to linear order in SOC~\cite{footnote1}. 
From Eq.~(\ref{eq_ksbe})  we derive coupled equations of motion for the inhomogeneous density and spin density of electrons and the density of holes. The spin density of the holes is assumed to be zero.  
\begin{figure}[bth]
\includegraphics[width=5.5cm]{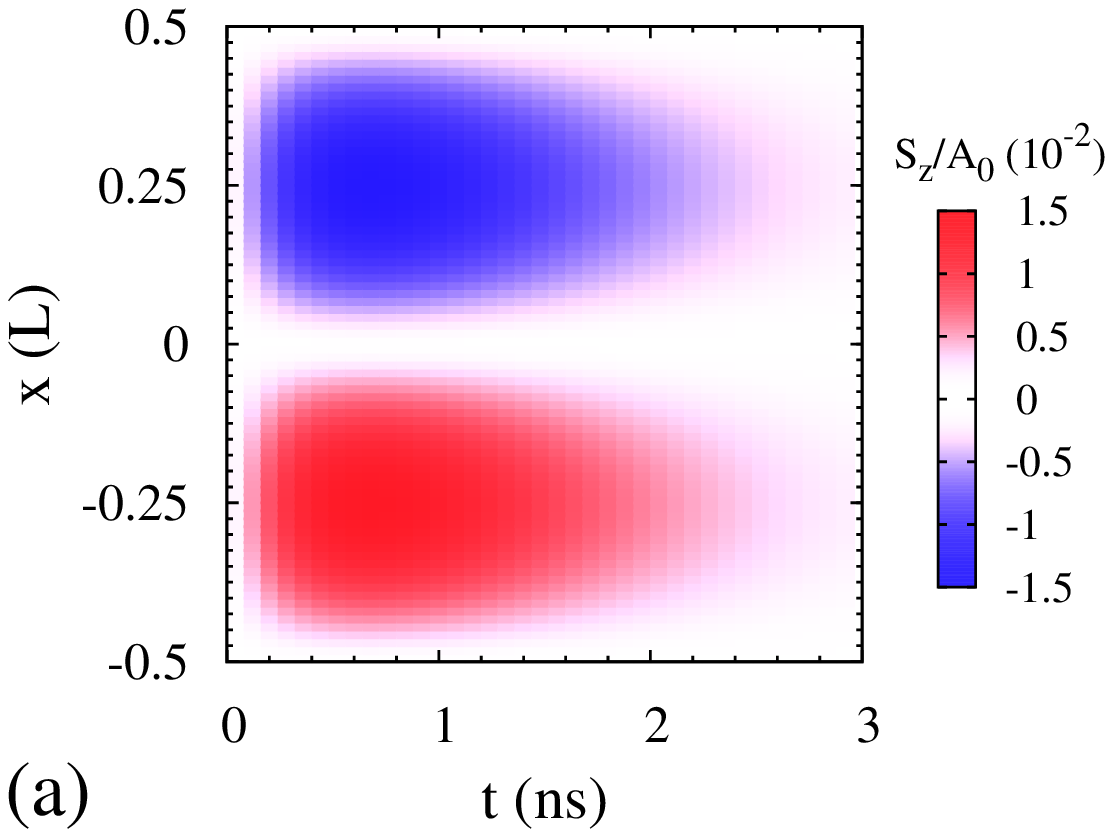}
\includegraphics[width=5.5cm]{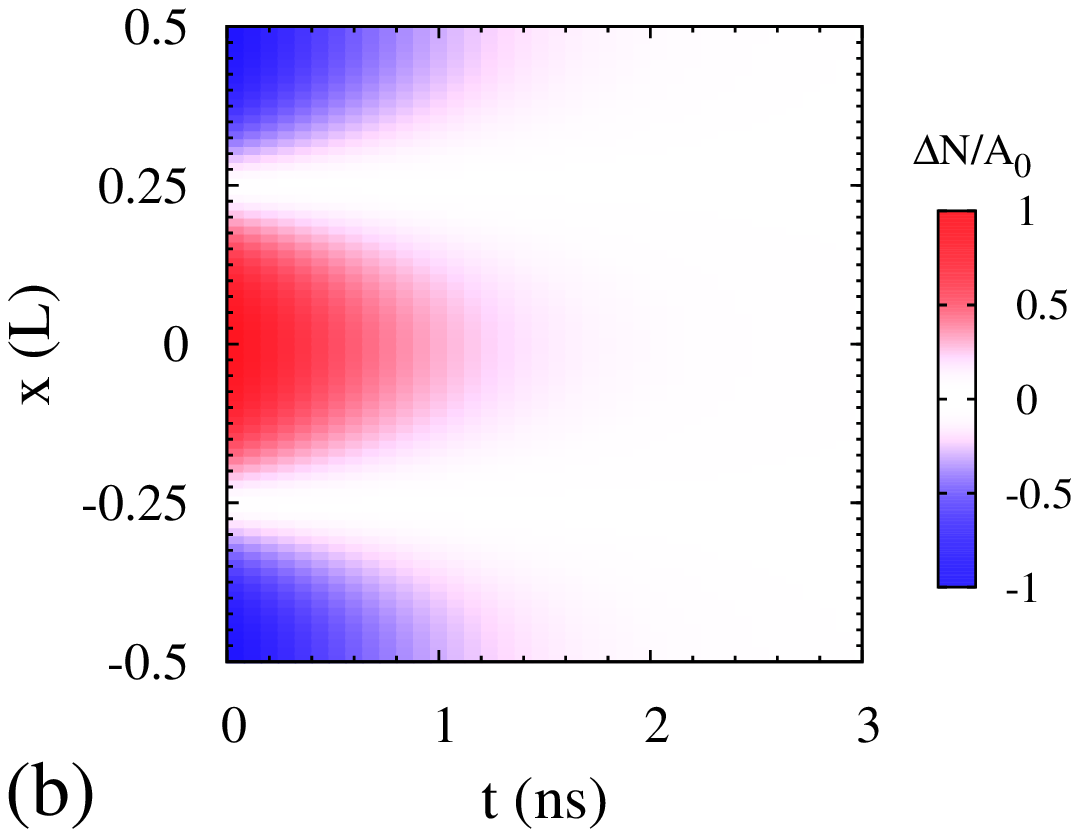}
\caption{(Color online) Time evolution of the electric-field-induced
  spin grating for electrons (a)
  (normalized by amplitude of the initial spin grating) as function of
  position (normalized by the wave length) with $q=0.3$~$\mu$m$^{-1}$
  in symmetric (110) QWs. (b) Time evolution of the
  e-h density grating. 
  Here, we take $E=1$~kV/cm, $D_a=20$~cm$^2$/s, $D_{s}=200$~cm$^{2}$/s,
  $\tau=1$~ps and $\Gamma=1$~ns$^{-1}$.
} 
\label{fig2}
\end{figure}

{\it (110) quantum well -- } For orientation, let us begin with the
simplest case, namely a symmetric  (110) GaAs QW.  What makes this
system  most interesting from our perspective is that the
Dresselhaus effective magnetic is along $z$-direction, and therefore
preserves the $z$-component of the electron spin, $S_z$.  The intrinsic SHE is completely absent.   The extrinsic SHE, embodied in the skew-scattering term, is present and clearly conserves $S_z$.  Therefore, we can write down separate kinetic equations for spin-up and spin-down electrons:
\begin{eqnarray}
  &&\hspace{-1cm}\partial_{t}n_{\sigma\mathbf{k}}+\frac{1}{m}\mathbf{k}\cdot\nabla_{\mathbf{r}}
  n_{\sigma\mathbf{k}}+e\mathbf{E}\cdot\nabla_{\mathbf{k}}n_{\sigma\mathbf{k}}
  \nonumber\\
  &&=-\frac{n_{\sigma\mathbf{k}}-n_{\sigma
      k}}{\tau}-\sigma\alpha_{ss}\sum_{|\mathbf{k}'|=|\mathbf
    k|}(\mathbf{k}\times\mathbf{k}'\cdot\hat{z})n_{\sigma 
    \mathbf{k}'},
\label{eq_18}\end{eqnarray}
with $\sigma=+,-$ representing  spin up and down with respect to
$z$-direction, respectively.
Following the standard procedure we substitute the ``first-order solution"
\begin{eqnarray}
n_{\sigma\mathbf{k}} &\approx& \bar n_{\sigma {\bf k}}
- \sigma\alpha_{ss} \sum_{|\mathbf{k}'|=|\mathbf
    k|}\left(\mathbf{k}\times\mathbf{k}'\cdot\hat{z}\right)  \bar n_{\sigma {\bf k'}}\,,
\end{eqnarray}
where $\bar n_{\sigma {\bf k}}\equiv \left(1-\frac{\tau}{m}\mathbf{k}\cdot\nabla_{\mathbf{r}} -e\tau\mathbf{E}\cdot\nabla_{\mathbf{k}}\right)n_{\sigma  k}$, 
into Eq.\,(\ref{eq_18}), and sum over ${\bf k}$ to obtain the diffusion equation 
\begin{equation}
\partial_{t}n_{\sigma}-D\nabla_{\mathbf{r}}^{2}n_{\sigma}+{\mathbf v}_d\cdot\nabla_{\mathbf{r}}n_{\sigma}
-\sigma \mathbf v_{ss}\cdot\nabla_{\mathbf{r}}n_{\sigma}=0,\end{equation}
where $n_\sigma=\sum_{\bf k}n_{\sigma\mathbf k}$ is the total
density of electron with spin $\sigma$ and $D=\langle
\frac{k^2}{2m^2}\tau\rangle$ is the diffusion constant. The
drift velocity and spin-Hall drift velocity are given by $\mathbf v_d
=\frac{\tau e\mathbf E}{m}$ and $\mathbf v_{ss}=2\alpha_{ss}\tau
eDm(\mathbf{E}\times\hat{z})$, respectively. We then combine the two
equations 
of different spins and get coupled kinetic equations for the total density ($N=n_++n_-$)
and the total spin polarization ($S_z=n_+-n_-$):
\begin{eqnarray}
(\partial_{t}-D\nabla_{\mathbf{r}}^{2}+\mathbf v_d\cdot\nabla_{\mathbf{r}})N
-\mathbf v_{ss}\cdot\nabla_{\mathbf{r}}S_z&=&0\,,\label{she1}\\
(\partial_{t}-D\nabla_{\mathbf{r}}^{2}+{\bf v}_d\cdot\nabla_{\mathbf{r}})S_z
-\mathbf v_{ss}\cdot\nabla_{\mathbf{r}}N&=&0\,.
\end{eqnarray}
Notice the appearance of a spin-density coupling, which occurs only in
a non-uniform system  and is proportional to the skew-scattering drift
velocity -- a quantity of first order in the SOC strength. 
The equation for the hole density is similar to Eq.\,(\ref{she1}), with $D$ and ${\bf v}_{d}$ replaced by the corresponding quantities for the holes,  but without the last term, because the spin polarization of the holes is neglected.  In fact, the last term can also be neglected on the left-hand side of Eq.\,(\ref{she1}) for the electrons, since it leads to minute corrections to the evolution of the density.
By imposing the local neutrality condition, that is,
assuming that the electron density is always equal to the
hole density, we combine the diffusion equations for electrons and holes 
into {\it ambipolar} diffusion and spin-density transport equations
\begin{eqnarray}
(\partial_{t}-D_a\nabla_{\mathbf{r}}^{2}+\Gamma)N &=&0\,,\\
(\partial_{t}-D_{s}\nabla_{\mathbf{r}}^{2})S_z
-\mathbf v_{ss}\cdot\nabla_{\mathbf{r}}N&=&0\,,
\end{eqnarray}
where $D_a$ and $D_{s}$ represent the ambipolar and spin diffusion
constants, respectively. 
Here, we have introduced the rate $\Gamma$ of electron-hole recombination. The solution of these equations is
\begin{eqnarray}
  \Delta N&=&A_0\cos(qx)e^{-(D_a q^2+\Gamma) t},\\
  S_z&=&-\tfrac{A_0\sin(qx)v_{ss}q}{(D_{s}-D_a)q^2-\Gamma}[e^{-(D_aq^2+\Gamma)t}-e^{-D_{s}q^2t}]\,.
  \label{110_dis}
\end{eqnarray}
In Fig.\,\ref{fig2}, we plot
the time evolution of the induced-spin grating as well as the density grating.
One can see that the amplitude of the spin
grating initially increases and then begins to decrease after a maximum
around 1\% the amplitude of the initial density grating. 
The induced spin grating shows a $\frac{\pi} 2$ phase shift from the density grating. 
From Eq.\,(\ref{110_dis}), we see that, for a given $q$,  $S_z$ reaches the maximal value
  \begin{equation}
\frac{A_{S_{z}}^{\rm
  max}(q)}{A_{0}}=\frac{v_{ss}q}{D_{a}q^{2}+\Gamma}\left(\frac{D_{s}q^{2}}{D_{a}q^{2}+\Gamma}\right)^{\frac{D_{s}q^{2}}{D_{a}q^{2}+\Gamma-D_{s}q^{2}}},\label{szmax}\end{equation}
 at  $t=({D_{a}q^{2}+\Gamma-D_{s}q^{2}})^{-1}\ln \left(\frac{D_{a}q^{2}+\Gamma}{D_{s}q^{2}}\right)$.  Noting that the quantity within the round brackets is of order $1$, we see that the amplitude ratio is roughly the fraction of the grating wavelength  covered by an electron that travels at the skew-scattering drift velocity ($v_{ss}$) during the diffusion lifetime of the grating ($1/D_aq^2$). 
Not surprisingly, this ratio  shows a non-monotonic dependence on
$q$, reaching a maximum $\frac{A_{S_{z}}^{\rm  max}(q^{\rm opt})}{A_0}\sim 1.4\times 10^{-2}$ at  the  optimal wave
vector $q^{\rm opt}\sim \unit{0.2}{\micro\meter^{-1}}$, with the material parameters listed
in the caption of Fig.\,\ref{fig2}.

{\it (001) quantum well -- } In a (001) QW, the presence of the in-plane effective magnetic field due to band SOC and the non-conservation of $S_z$   lead to  more complex scenarios.  To begin with, the  coupling of longitudinal and transverse spin fluctuations leads to a set of  drift-diffusion equations of the form 
\begin{equation}\label{DDE}
\partial_t (\Delta N,S_x,S_y,S_z)^T=-{\cal D}(\tilde {\mathbf q})(\Delta N,S_x,S_y,S_z)^T\,,
\end{equation}
\\where ${\cal D}(\tilde {\mathbf q})$ is the $4\times 4$
drift-diffusion matrix acting on the column vector of the Fourier
amplitudes of the density at wave vector ${\bf q}$.    Here
$\tilde{\bf q}:\equiv{\bf q}-ie{\bf E}\partial_{\epsilon_k}$ is
a momentum-space operator, which takes into account drift under the
action of the electric field ${\bf E}$.  Without going into technical
details we only summarize the salient results  (for details, see Ref.~\onlinecite{SMF}).  Taking ${\bf q}=q {\bf
  \hat x}$ and ${\bf E}=E{\bf \hat y}$ and assuming
$\frac{k_Fq}{m}\ll1$ and $|\alpha\pm\beta|k_F \ll E_F$
(conditions that define the diffusive regime) we find
\begin{widetext}
\begin{equation}
{\cal D}(\tilde {\mathbf q}) = 
 \left(\begin{array}{cccc}
Dq^2 & -\frac{1}{2}\tau \lambda_2
    q^2v_d & \begin{array}{c}-4i\tau
D\lambda_1\tilde\lambda_2^2q+i\tau\tilde\lambda_2\gamma_yqv_d \\
\mbox{}-iqv_{ss}D\tilde\lambda_2/v_d\end{array}
& \begin{array}{c}-4i\tau D\tilde\lambda_2^2\gamma_y q-i\tau\lambda_1\tilde\lambda_2qv_d\\-iqv_{ss}\end{array}\\
\begin{array}{c}-\frac{1}{2}\tau \lambda_2
q^2v_d+4\tau \lambda_2
(\tilde \lambda_1^2+\tilde\gamma_y^2)v_d\\\mbox{}+ 2\tilde\lambda_1v_{ss}\end{array}
 & Dq^2+\frac{1}{\tau_{sx}} & 
-4i D\tilde{\gamma_{y}} q & 4i
D\tilde{\lambda_{1}} q\\
\begin{array}{c}-4i\tau D\lambda_1\tilde\lambda_2^2
  q+i\tau\tilde\lambda_2\gamma_yqv_d\\
-iqv_{ss}D\tilde\lambda_2/v_d
\end{array} & 4i D\tilde{\gamma_{y}}q 
& Dq^2+\frac{1}{\tau_{sy}} &
-4 D\tilde{\lambda_{1}}\tilde{\gamma_{y}}+2
\tilde{\lambda_{2}} v_d \\
\begin{array}{c}-4i\tau D \tilde\lambda_2^2\gamma_y q -i\tau\lambda_1\tilde\lambda_2qv_d\\\mbox{}-iqv_{ss}\end{array}
& -4i D\tilde{\lambda_{1}} q & -4
D\tilde{\lambda_{1}}\tilde{\gamma_{y}}-2
\tilde{\lambda_{2}}v_d &
Dq^2+\frac{1}{\tau_{sz}}\end{array}\right)
\label{D_k}
\end{equation}
\end{widetext}
where $v_d=\frac{\tau e E}{m}$.
Here, $\tilde{\lambda}_{i}=m\lambda_{i}$, $\tilde{\gamma_{i}}=m\gamma_{i}$, 
$\frac{1}{\tau_{sx}}=4D(\tilde\lambda_{1}^{2}+\tilde\gamma_{y}^{2})$, 
$\frac{1}{\tau_{sy}}=4D(\tilde\lambda_{2}^{2}+\tilde\gamma_{y}^{2})$ and
$\frac{1}{\tau_{sz}}=4D(\tilde\lambda_{1}^{2}+\tilde\lambda_{2}^{2})$.

We note that our diffusion matrix differs from the one
reported in Ref.\,\onlinecite{Anderson10} in two ways:  (i) in addition to the ``standard" terms linear in  $\tilde {\bf q}$ and cubic in the SOC strength, we include
terms of second order in both
$\tilde{\bf q}$  and  the SOC strength as well as terms of third order in $\tilde{\bf q}$ and first order in SOC.  All these terms can be of comparable magnitude in real systems. 
(ii) At variance with Ref.\,\onlinecite{Anderson10}, our diffusion matrix  is non-symmetric:  ${\cal D}_{i1}\neq {\cal D}_{1i}$.  This lack of symmetry comes from a careful consideration of the operatorial character of ${\bf \tilde q}$, whereby ${\bf \tilde q}\epsilon_k \neq \epsilon_k{\bf \tilde q}$, as explained in the supplemental material~\cite{SMF}.  Eqs.~(\ref{DDE}-\ref{D_k}) are our main theoretical result: they combine extrinsic and intrinsic contributions to the SHE as well as spin precession, and reduce to the results of the previous section if the intrinsic SOCs appropriate for (110) QW are used.

{\it (001) quantum well with balanced SOC -- } The case of a (001) QWs with identical Dresselhaus and Rashba coefficients, $\alpha=\beta$  (corresponding to the condition $\lambda_2=0$)  with ${\bf q}$ oriented along the [110] direction gives us the opportunity to demonstrate a particularly interesting application of Eqs.~(\ref{DDE}-\ref{D_k}).   
Just as in a symmetric (110) QW, only the skew scattering contributes to the collective SHE, but now $S_z$ is not conserved.  Since  $\gamma_y$ is negligibly small (two orders smaller than the band SOC), the  $S_y$
component decouples from  the $S_x$ and $S_z$ components and the diffusion matrix  reduces to 
\begin{equation}
{\cal D}(q) = 
 \left(\begin{array}{cccc}
Dq^2 & 0 & 0
& -iqv_{ss}  \\
q_0 v_{ss}& D(q^2+q_0^2) & 
0
& 2iDqq_0\\
0 & 0
& Dq^2 &0\\
-iqv_{ss} & -2iDqq_0 & 0&
D(q^2+q_0^2)\end{array}\right),
\end{equation}
with $q_0=\frac{4m\beta}{\hbar^2} \simeq 3.5~\mu$m$^{-1}$.  After
imposing the charge-neutrality condition, the diffusion
equations for the density and the two helical components of the spin density
$S_\pm=\frac{1}{\sqrt{2}}(S_x\pm iS_z)$ are found to be 
\begin{eqnarray}
  \partial_t \Delta N&=&-D_aq^2\Delta N-\tfrac{1}{\sqrt 2}
  qv_{ss}S_-+\tfrac{1}{\sqrt 2}qv_{ss}S_+,\label{eq24}\\
  \partial_t S_-&=&\tfrac{1}{\sqrt 2}(q+q_0)v_{ss}N-D_s(q+q_0)^2S_-,\\
  \partial_t S_+&=&-\tfrac{1}{\sqrt 2}(q-q_0)v_{ss}N-D_s(q-q_0)^2S_+.\label{eq26}
\end{eqnarray}
As in the previous calculations, we neglect the last two terms on
the right-hand side of Eq.\,(\ref{eq24}).  Then the solution for the density reduces to a simple diffusion process, and the solution for the
two helical modes is given by
\begin{equation}
  S_{\mp}=\tfrac{\pm \frac{1}{\sqrt 2}q_\pm
    A_0e^{iqx}v_{ss}}{D_{s}q_\pm^2-D_aq^2}[e^{-(D_aq^2-D_{s}q_\pm^2)
    t}-1]e^{-D_{s}q_\pm^2t},\label{twomodes}
\end{equation}
which yields the spin-polarization
\begin{eqnarray}
  S_{x}&=&\sum_\pm\tfrac{\pm \frac{1}{2}q_\pm
    A_0\cos(qx)v_{ss}}{D_{s}q_\pm^2-D_aq^2}[e^{-D_aq^2t}-e^{-D_{s}q_\pm^2t}],\\
  S_{z}&=&-\sum_\pm\tfrac{\frac{1}{2}q_\pm
    A_0\sin(qx)v_{ss}}{D_{s}q_\pm^2-D_aq^2}[e^{-D_aq^2t}-e^{-D_{s}q_\pm^2t}].
\end{eqnarray}
These amplitudes show a strong dependence on the wave vector. 
One can see that the contributions from
 the  $S_-$ mode is proportional to $q_+=q+ q_0$ while the contribution 
 from the  $S_+$ mode is proportional to $q_-=q- q_0$ (the correspondence is reversed if we switch the sign of $\beta$).  Further, the $S_+$ mode is long-lived, due to the slowly decaying term $e^{-D_{s}q_-^2t}$, while the $S_-$ mode is short-lived~\cite{Bernevig06,Weber07,Weng08,Yang12,Tokatly13}.
The long-time behavior of $S_z$, being dominated by the $S_+$ component, is positive for $q>q_0$ and negative for $q<q_0$.  In the special case $q=q_0$ -- a practically realizable case -- $S_+$ vanishes identically, and the amplitude of $S_z$ decays to zero most rapidly.  In this case, the skew scattering converts  the initial density grating into a helical wave of wave vector $q_0$!  Further interpretation of this intriguing effect, based on an SU(2) gauge transformation that eliminates the intrinsic SOC~\cite{Bernevig06,Tokatly13}, is given in Ref.~\onlinecite{SMF}.   

\begin{figure}[]
  \includegraphics[width=6cm]{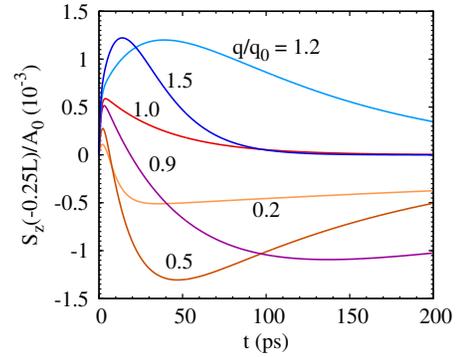}
\caption{(Color online) Time evolution of the spin component,
  $S_z$, from the density grating with different values of $q/q_0$ in
  the case of $\alpha=\beta$, $q_0=\frac{4 m\beta}{\hbar^2}$. In the
  calculation, we take the Dresselhaus coefficient
  $\beta=10$~meV\AA (corresponding to 10~nm GaAs QW) and $q_0 \simeq 3.5~\mu$m$^{-1}$. Other parameters
  are taken to be the same as Fig.\,\ref{fig2}.} 
\label{fig3}
\end{figure}

In Fig.~\ref{fig3}, we  plot the amplitude of the $z$-component of the electron spin, $S_z$,
as function of time  at a distance $x=-0.25 L$  from a peak of the
density grating. Notice the reversal of sign of the long-time behavior and the quick decay of the signal at $q=q_0$, due to the vanishing of the $S_+$ mode. 

In summary, we have studied the collective spin Hall effect in a periodically
modulated electron gas in the presence of an in-plane electric field perpendicular to the wave vector of the initial density modulation.
In the symmetric (110) quantum well  the amplitude of the induced spin density is controlled solely by skew scattering and can be as large as 1\% of that of
the initial density modulation.   This should be observable in state-of-the art experiments~\cite{Cameron96,weber_nature_2005,Yang12}. Similarly, the collective spin Hall effect
in (001) QWs with identical Rashba and Dresselhaus SOC strengths is
also entirely controlled  by  skew scattering.  
In this case, the skew scattering generates a spiral spin density wave when the wave vector of the initial grating matches the wave vector of the spin-orbit coupling.

We gratefully acknowledge support  from NSF Grant No. DMR-1104788.


\begin{thebibliography}{100}
\bibitem{footnote0}That is, if only the spin-density coupling that is linear in the wave vector $q$ of the grating is considered.
\bibitem{footnotex}{We have temporarily reinstated $\hbar$ to point out that $\alpha_{ss}$ has the dimensions of a diffusion constant}
\bibitem{footnote1}We point out that the correction due to SOC
with the in-plane electric field accounts for only one half of the so-called side-jump
contribution to the extrinsic spin Hall effect~\cite{Culcer10}.
\end{thebibliography}
\end{document}


\title{Supplemental material file for ``Collective spin-Hall effect for electron-hole gratings"}
\author{Ka Shen and G. Vignale}
\affiliation{Department of Physics and Astronomy, University of
Missouri, Columbia, Missouri 65211, USA} 

\maketitle
\makeatletter 
\makeatother

This  material includes the detailed derivation of the drift-diffusion
equation in (001) GaAs [Eqs. (14) and (15) of main text] and further discussions of its structure, in particular the asymmetry of the diffusion matrix.  We also supply the details of the  SU(2) transformation of the skew scattering for the case of identical Rashba and
Dresselhaus coupling strengths.

\section{1. Derivation of drift-diffusion equation in (001) GaAs quantum wells}
 By defining the $x$
and $y$ axes in the [110] and [$\bar 1$10] direction separately, we
express the Hamiltonian of
electron gas in (001) GaAs QWs as
\begin{equation}
H_{0}=\frac{k^{2}}{2m_e}+ k_x(\lambda_1\sigma_{y}+\gamma_y\sigma_z)+k_y(\lambda_2\sigma_{x}-\gamma_x\sigma_z),
\label{hamiltonian}
\end{equation}
where $\lambda_1=\beta+\alpha$ and $\lambda_2=\beta-\alpha$ with
$\alpha$ and $\beta$ being the Rashba and Dresselhaus
coefficients. $\gamma_i=\frac{\lambda_{e}^2}{4} eE_i$ describes the
SOC due to the in-plane electric field with $\lambda_{e}$
corresponding to the effective Compton wavelength.
In the relaxation time approximation, the kinetic equation for the local
electron density matrix $\rho_{\bf  k}({\bf r})$ is given by~\cite{Mishchenko04,Bernevig06}
\begin{eqnarray}
&&\partial_{t}\rho_{\mathbf{k}}+\frac{1}{2}\{\nabla_{\mathbf{k}}H_{0},\tilde\nabla_{\mathbf{r}}\rho_{\mathbf{k}}\}
+i[H_{0},\rho_{\mathbf{k}}]\nonumber\\
&=&-\frac{\rho_{\mathbf{k}}}{\tau}+\frac{\rho_{k}}
  {\tau}+\frac{1}{2m\tau}\{\mathbf{k}\cdot\mathbf{A},\partial_{\epsilon_{k}}\rho_{k}\}
  \nonumber\\
&&\mbox{}-\frac{1}{2}\alpha_{ss} \sum_{|\mathbf{k}'|=|\mathbf k|}
\{\mathbf{k}\times\mathbf{k}'\cdot\bg\sigma,\rho_{\mathbf{k}'}\},\label{eq_ksbe}\end{eqnarray} 
where $\tilde \nabla_{\mathbf{r}}=\nabla_{\bf r}+e{\bf
  E}\partial_{\epsilon_k}$. 
The last term on the right-hand side is the skew scattering term from
the second-order Born approximation~\cite{Cheng08,Raimondi12}. 
With Fourier transform with respect to
$t$ and $\mathbf r$, we rewrite Eq.\,(\ref{eq_ksbe}) as~\cite{Liu12}
\begin{equation}
  ({\cal I}+{\cal K}_{\bf k})\left(\begin{array}{c}
      g_{\bf k}^{0}\\
      g_{\bf k}^x\\
      g_{\bf k}^{y}\\
      g_{\bf k}^{z}\end{array}\right)=({\cal I}+{\cal T}_{\bf k})\left(\begin{array}{c}
      g_k^{0}\\
      g_k^{x}\\
      g_k^{y}\\
      g_k^{z}\end{array}\right)+{\cal M}_{\mathbf k,\mathbf k'}\left(\begin{array}{c}
      g_{\bf k'}^{0}\\
      g_{\bf k'}^x\\
      g_{\bf k'}^{y}\\
      g_{\bf k'}^{z}\end{array}\right),\label{eq_KT}
\end{equation}
with $\rho_{\bf k}=\sum_i g_{\bf k}^i \sigma_i$ and $\rho_{
  k}=\sum_i g_k^i \sigma_i$. Here ${\cal I}$ represents the $4\times 4$
identical matrix, and ${\cal T}$  and ${\cal K}$
are given by
\begin{equation}
{\cal T}_{\bf k}=\left(\begin{array}{cccc}
    0 & {-B_x}\partial_{\epsilon_{k}} & -B_y\partial_{\epsilon_{k}} & -B_z\partial_{\epsilon_{k}}\\
{-B_x}\partial_{\epsilon_{k}} & 0 & 0 & 0\\
{-B_y}\partial_{\epsilon_{k}} & 0 & 0 & 0\\
{-B_z}\partial_{\epsilon_{k}} & 0 & 0 &
0\end{array}\right),\end{equation} 
and \begin{eqnarray}
&&\hspace{-0.5cm}{\cal K}_{\bf k}=\nonumber\\
&&\hspace{-0.5cm}\left(\begin{array}{cccc}
\Omega & i\lambda_{2}\tau \tilde q_{y} & i\lambda_{1}\tau \tilde q_{x}
& i\tau(\gamma_{y}\tilde q_{x}-\gamma_{x} \tilde q_{y})\\
i\lambda_{2}\tau \tilde q_{y} & \Omega & -2B_{z}\tau & 2B_{y}\tau\\
i\lambda_{1}\tau \tilde q_{x} & 2B_{z}\tau & \Omega & -2B_{x}\tau\\
i\tau(\gamma_{y}\tilde q_{x}-\gamma_{x} \tilde q_{y}) & -2B_{y}\tau & 2B_{x}\tau &
\Omega\end{array}\right),\nonumber\\
\end{eqnarray}
respectively, where $B_{x}  =  -k_{y}\lambda_{2}$,
$B_{y} = -k_{x}\lambda_{1}$,
$B_{z} = -(k_{x}\gamma_{y}-k_{y}\gamma_{x})$, and
$\Omega=-i\omega\tau+i\frac{\tau}{m}\mathbf{k}\cdot\tilde{\mathbf{q}}$.
The effective wave vector is given by $\tilde{\bf q}={\bf q}-ie{\bf
  E}\partial_{\epsilon_k}$. The matrix, ${\cal M}$, describing skew
scattering, is given by
\begin{equation}
{\cal M}_{\mathbf k,\mathbf k'}=-\alpha_{ss}\tau \left(\begin{array}{cccc}
    0 & 0 & 0 & k_x k_y'-k_yk_x'\\
    0 & 0 & 0 & 0\\
    0 & 0 & 0 & 0\\
    k_x k_y'-k_yk_x' & 0 & 0 &
0\end{array}\right).\end{equation}
By multiplying $({\cal I}+{\cal K}_{\bf k})^{-1}$ to
Eq.\,(\ref{eq_KT}) and substituting the $g_{\bf k}^i=\sum_j[({\cal I}+{\cal
  K}_{\bf k})^{-1}({\cal I}+{\cal T}_{\bf k})]_{ij}g_{k}^j$ into the
skew scattering term, we obtain
\begin{equation}
[{\cal I}-\langle ({\cal I}+\frac{1}{{\cal I}+{\cal K}_{\bf k'}}{\cal
  M}_{\mathbf k',\mathbf k} ) \frac{1}{{\cal
  I}+{\cal K}_{\bf k}}({\cal I+T}_{\bf
  k})\rangle]\left(\begin{array}{c} 
g_k^{0}\\
g_k^{x}\\
g_k^{y}\\
g_k^{z}\end{array}\right)=0.\end{equation}
Here, $\langle...\rangle$ represents the average over the angular
dependence of the momentum.
By further averaging over the distribution function with respect to
energy,  retaining $\omega$ up to the first order
under the condition $\omega\tau\ll 1$, and Fourier transforming
back to the time domain we get the final  diffusion equation in the form 
\begin{equation}\label{DDE}
\partial_t (\Delta N,S_x,S_y,S_z)^T=-{\cal D}(\tilde {\mathbf q})(\Delta N,S_x,S_y,S_z)^T\,.
\end{equation}
with
\begin{equation}
{\cal D} (\tilde{\mathbf q})={\cal I}-\langle ({\cal I}+\frac{1}{{\cal I}+{\cal K}_{\bf k'}}{\cal
  M}_{\mathbf k',\mathbf k} ) \frac{1}{{\cal
  I}+{\cal K}_{\bf k}}({\cal I+T}_{\bf
  k})\rangle|_{\omega=0}.
\end{equation}
For $\frac{k_Fq}{m}$ and $|\alpha\pm\beta|k_F \ll E_F$, we do perturbation expansion with respect to
${\cal T}$ and ${\cal K}$. By substituting $\tilde{\bf q}:=q\hat x-ieE\hat
y\partial_{\epsilon_k}$, we finally obtain the diffusion matrix,
written up to the third order of ${\cal K}$, (including ${\cal
T}_{\bf k}$, ${\cal K}_{\bf k}$, ${\cal K}_{\bf k}{\cal T}_{\bf
k}$,${\cal K}_{\bf k}^2$, ${\cal K}^2_{\bf k}{\cal T}_{\bf k}$, ${\cal
K}^3_{\bf k}$, ${\cal K}^3_{\bf k}{\cal T}_{\bf k}$, ${\cal K}_{\mathbf
k'}{\cal M}_{\mathbf k',\mathbf k}{\cal K}_{\mathbf k}$, ${\cal K}_{\mathbf
k'}{\cal M}_{\mathbf k',\mathbf k}{\cal T}_{\mathbf k}$):
\begin{widetext}
\begin{equation}
{\cal D}(\tilde {\mathbf q}) = 
 \left(\begin{array}{cccc}
Dq^2 & -\frac{1}{2}\tau \lambda_2
    q^2v_d & \begin{array}{c}-4i\tau
D\lambda_1\tilde\lambda_2^2q+i\tau\tilde\lambda_2\gamma_yqv_d \\
\mbox{}-iqv_{ss}D\tilde\lambda_2/v_d\end{array}
& \begin{array}{c}-4i\tau D\tilde\lambda_2^2\gamma_y q-i\tau\lambda_1\tilde\lambda_2qv_d\\-iqv_{ss}\end{array}\\
\begin{array}{c}-\frac{1}{2}\tau \lambda_2
q^2v_d+4\tau \lambda_2
(\tilde \lambda_1^2+\tilde\gamma_y^2)v_d\\\mbox{}+ 2\tilde\lambda_1v_{ss}\end{array}
 & Dq^2+\frac{1}{\tau_{sx}} & 
-4i D\tilde{\gamma_{y}} q & 4i
D\tilde{\lambda_{1}} q\\
\begin{array}{c}-4i\tau D\lambda_1\tilde\lambda_2^2
  q+i\tau\tilde\lambda_2\gamma_yqv_d\\
-iqv_{ss}D\tilde\lambda_2/v_d
\end{array} & 4i D\tilde{\gamma_{y}}q 
& Dq^2+\frac{1}{\tau_{sy}} &
-4 D\tilde{\lambda_{1}}\tilde{\gamma_{y}}+2
\tilde{\lambda_{2}} v_d \\
\begin{array}{c}-4i\tau D \tilde\lambda_2^2\gamma_y q -i\tau\lambda_1\tilde\lambda_2qv_d\\\mbox{}-iqv_{ss}\end{array}
& -4i D\tilde{\lambda_{1}} q & -4
D\tilde{\lambda_{1}}\tilde{\gamma_{y}}-2
\tilde{\lambda_{2}}v_d &
Dq^2+\frac{1}{\tau_{sz}}\end{array}\right),\label{D_k}
\end{equation}
\end{widetext}
with $v_d=\frac{\tau e E}{m}$ and $ v_{ss}=2\alpha_{ss}\tau
eDmE$.
Here, $\tilde{\lambda}_{i}=m\lambda_{i}$, $\tilde{\gamma_{i}}=m\gamma_{i}$, 
$\frac{1}{\tau_{sx}}=4D(\tilde\lambda_{1}^{2}+\tilde\gamma_{y}^{2})$, 
$\frac{1}{\tau_{sy}}=4D(\tilde\lambda_{2}^{2}+\tilde\gamma_{y}^{2})$ and
$\frac{1}{\tau_{sz}}=4D(\tilde\lambda_{1}^{2}+\tilde\lambda_{2}^{2})$.

\section{2. Asymmetry of the diffusion matrix}
The above diffusion matrix, Eq.\,(\ref{D_k}) [i.e., Eq.\,(15) in
main text], clearly shows that the
spin-charge coupling is non-symmetric, i.e., ${\cal D}_{1i}\ne{\cal
  D}_{i1}$. This comes from
our careful consideration of the order of the quantity $\epsilon_k$
and the operator $\partial_{\epsilon_k}$ in $\tilde q_y$, e.g.,
\begin{eqnarray}
\langle \tilde q_y \epsilon_k \rangle &=& -i\frac{1}{N} e
E \sum_{\bf  k} \partial_{\epsilon_{k}} (\epsilon_k\rho_k)=0,\\
\langle \epsilon_k \tilde q_y \rangle &=& -i\frac{1}{N} e{ E}\sum_{\bf
  k} \epsilon_k \partial_{\epsilon_{k}} \rho_k=i\frac{m}{\tau}{ v}_d.
\end{eqnarray}
Notice that the asymmetry here is created by the external electric field: no violation of Onsager's reciprocity relations is implied.  
If one erroneously assumed $\langle \tilde q_y \epsilon_k
\rangle=\langle \epsilon_k \tilde q_y \rangle\ne 0$, as was apparently done in
Ref.\,\onlinecite{Anderson10}, the diffusion matrix would be
symmetric and both spin-density and density-spin couplings would differ from zero  in a homogeneous system ($q=0$). 
To better understand the implications of this fact, observe that in the homogeneous case
our diffusion equation reduces to
\begin{widetext}
\begin{equation}\label{DDE}
\partial_t \left(\begin{array}{c} N\\S_x\\S_y\\S_z\end{array}\right)
=- \left(\begin{array}{cccc}
0 & 0 & 0 & 0\\
\begin{array}{c}4\tau \lambda_2
(\tilde \lambda_1^2+\tilde\gamma_y^2)v_d+ 2\tilde\lambda_1v_{ss}\end{array}
 &\frac{1}{\tau_{sx}} & 
0 & 0\\
0 & 0
& \frac{1}{\tau_{sy}} &
-4 D\tilde{\lambda_{1}}\tilde{\gamma_{y}}+2
\tilde{\lambda_{2}} v_d \\
0& 0 & -4D\tilde{\lambda_{1}}\tilde{\gamma_{y}}-2
\tilde{\lambda_{2}}v_d &
\frac{1}{\tau_{sz}}\end{array}\right)\left(\begin{array}{c} N\\S_x\\S_y\\S_z\end{array}\right)\,.
\end{equation}
\end{widetext}
Obviously, the complete vanishing of the ${\cal D}_{1i}$ guarantees
the conservation of the particle number (i.e., the $q=0$ component of the particle density)  -- an exact physical constraint.   However,  a non-zero ${\cal D}_{12}$ arising from the incorrect treatment of the ordering of the operators would violate this constraint when a uniform spin polarization is present.
On the other hand, the non-zero matrix element ${\cal D}_{21}$
in Eq.\,(\ref{DDE})
predicts the generation of a uniform in-plane spin polarization by a steady electric current, 
in the presence of SOC.   This is a well-known
effect -- the so-called Edelstein effect~\cite{Aronov89,Edelstein90} --  and has been observed in
experiments~\cite{Silsbee04}.
 For example, for a pure Rashba SOC ($\tilde \lambda_1=-\tilde
  \lambda_2= m\alpha$) in the steady state we obtain, to linear order
  in the electric field, 
  \begin{equation}\label{Edelstein}
P=\frac{S_x}{N}=\frac{\alpha\tau}{D}v_d-\frac{v_{ss}}{2D\alpha
  m}\,.
  \end{equation}
  The first term is
solely due to Rashba SOC~\cite{Aronov89,Edelstein90,Silsbee04}, while
the second term describes the 
correction due to the skew scattering~\cite{Cheng08,Raimondi12}.  
We should point out that in our derivation so far 
the spin relaxation has been assumed to be dominated by the 
D'yakonov-Perel'  (DP) mechanism: therefore, $\alpha$ should be finite. In the limit 
$\alpha\to 0$,  different  spin relaxation mechanisms, e.g., the Elliott-Yafet (EY) 
mechanism, become dominant. 
 By replacing $\frac{1}{\tau_s}$ by $\frac{1}{\tau_s^{\rm DP}}+\frac{1}{\tau_s^{\rm EY}}$ in our theory, we see that the spin polarization  
vanishes in the $\alpha \to 0$ limit as physically expected, and  in agreement with previous work~\cite{Raimondi12}.

\section{3. Analysis of the balanced case: $\alpha=\beta$}
To better understand the generation of helical spin modes by
skew scattering in the case of balanced SOC ($\alpha=\beta$) in a (001) GaAs quantum well we recall that in this special case the SOC can be completely eliminated by an SU(2) gauge transformation, which is actually a non-uniform rotation in spin space~\cite{Bernevig06,Tokatly13}.  The spin dynamics in the rotated reference frame is simply determined by the competition of the  spin-conserving drift-diffusion process and the gauge-transformed skew-scattering process.  The  gauge-transformed skew-scattering term has the form
\begin{eqnarray}
\partial_t\tilde{\rho}_{\bf k}|_{\rm scat}^{ss}&=&
-\frac{\alpha_{ss}}{2} \sum_{|\mathbf{k}'|=|\mathbf k|}
(\mathbf{k}\times\mathbf{k}')_z\{\cos(q_0x)\sigma_z,{\tilde
  \rho}_{\mathbf{k}'}\}\nonumber\\
&&\hspace{1cm}\mbox{}+(\mathbf{k}\times\mathbf{k}')_z\{\sin(q_0x)\sigma_x,{\tilde \rho}_{\mathbf{k}'}\},
\end{eqnarray}
which suggests that the spin Hall velocity in the rotated frame has a sinusoidal variation in space, with the wave vector $q_0=\frac{4m\beta}{\hbar^2}$. More precisely, the
transverse drift velocity for the $\tilde {S}_z$ component is 
$v_{ss}^z(x)=v_{ss}\cos(q_0x)$, while that of  the $\tilde{S_x}$  component is
$v_{ss}^x(x)=v_{ss}\sin(q_0x)$. Since the
band SOC is completely gauged away, one has the carrier conservation equation
\begin{equation}
  \frac{d\tilde{\rho}^{z(x)}_\sigma(x,t)}{dt}=\partial_t\tilde{\rho}^{z(x)}_\sigma(x,t)
  -\sigma\partial_x[\tilde{\rho}^{z(x)}_\sigma(x,t)v^{z(x)}_{ss}(x)]\equiv 0\,,
\end{equation}
where the diffusion process has been neglected for simplicity.
By substituting
$\tilde{\rho}^{z(x)}_\sigma(x,t)\simeq \frac{1}{2}A_0
\cos(qx)$, we obtain the equations for ${\tilde S}_z$ and ${\tilde
  S}_x$ in the following form:
$$\partial_t\tilde{S_z}(x)=-\frac{1}{2}A_0v_{ss}[q_-\sin(q_-x)+q_+\sin(q_+x)]\,,$$
and
$$\partial_t\tilde{S_x}(x)=-\frac{1}{2}A_0v_{ss}[q_-\cos(q_-x)-q_+\cos(q_+x)]\,.$$
The component with wave vector $q_-$ 
corresponds to the slowly decaying  helical mode $S_+$, whose amplitude is  proportional to $q_-$.   Similarly, the component with wave vector $q_+$ 
corresponds to the rapidly decaying  helical mode $S_-$,  with
amplitude proportional to $q_+$.  
By transforming back to the original, unrotated spin space, we  obtain 
\begin{eqnarray}
 \partial_t{S_x}&=&\cos(q_0x)\partial_t\tilde S_x-\sin(q_0x)\partial_t\tilde
  S_z\nonumber\\
  &=&\sum_\pm\pm\frac{1}{2}q_\pm A_0\cos(qx)v_{ss}\,,\\
  \partial_t{S_z}&=&\sin(q_0x)\partial_t\tilde S_x+\cos(q_0x)\partial_t\tilde S_z\nonumber\\
  &=&-\sum_\pm\frac{1}{2}q_\pm A_0\sin(qx)v_{ss}\,,
\end{eqnarray}
which are consistent with Eqs.\,(21-22) in main text. From this analysis one
can see that in the SU(2)-rotated frame the skew scattering cannot create a uniform spin
polarization,  which would correspond to the $S_+$ mode in the original frame.  This is the reason for the vanishing amplitude of the ${\tilde S}_+$ mode at $q=q_0$.